\documentclass[prb,article]{revtex4}
\usepackage{graphicx}
\usepackage{amsmath}
\usepackage{amsfonts}
\usepackage{amssymb}
\providecommand{\U}[1]{\protect\rule{.1in}{.1in}}

\begin{document}

\title{Crystal mean field based trial wave functions for the FQHE ground states}
\author{Alejandro Cabo$^{\ast}$ and Francisco Claro$^{\ast\ast}$}

\affiliation{$^{*}$ Grupo de F\'{\i}%
sica Te\'orica, Instituto de Cibern\'etica, Matem\'atica y
F\'{\i}sica, Calle E, No. 309, Vedado, La Habana, Cuba}
\affiliation
{$^{**}$ Facultad de F\'{\i}%
sica, Pontificia Universidad Cat\'olica de Chile, Vicu\~na
Mackenna 4860, 6904411 Macul, Santiago, Chile}

\begin{abstract}
\noindent Employing the Haldane-Rezayi periodic representation,
the crystalline determinantal Hall crystal mean field solutions
derived in previous works are used to construct variational wave
functions for the FQHE at $\nu=1/q$. The proposed states optimize
the short range correlations in a similar measure as the Laughlin
ones, since the zero of the states when the coordinates of two
particles join is of order $q$. However, the proposed wave
functions also incorporate the crystalline correlations of the
mean field problem, through a determinantal mean field function
entering their construction. The above properties, lead to the
expectation that the considered states can be competitive in
energy per particle with the Laughlin ones. Their similar
structure also could explain way the breaking of the translation
invariance in the FQHE ground states can result to be a weak one,
which after disregarded, produce the Laughlin states as good
approximations. Calculation for checking these possibilities are
under consideration.

\medskip


\ \bigskip
\end{abstract}
\maketitle

%

\section{Introduction}

In a \ previous work (See Ref. [\onlinecite{ccpm}]) analytical
solutions of the Hartree-Fock problem in a magnetic field
corresponding to filling factors $\nu=1/q$ \ ($q$ odd) and a
number of particles per cell $\gamma=1,1/2$, \ were determined.
The form of the obtained wave functions directly suggested a
possible way of constructing \ variational states, being able to
reduce the energy through optimizing both: short range as well as
crystalline correlations. The appropriate picture for this purpose
appeared to be the procedure to impose periodic boundary
conditions in a magnetic field developed by Haldane and Rezayi in
Ref. [\onlinecite{halrez}]. \

\ \ \ This paper intends to construct the mentioned states. \ The
main idea is the following: the mean field one particle states
after written in the Haldane-Rezayi picture, have a structure
which is given by the product of a determinantal function which
contains all the dependence of the quantum numbers of the HF
states, times a factor which have a large number of zeroes which
are completely attached to the periodicity lattice of the HF
states. The position of those zeroes have no dependence of the set
of the quantum numbers of the filled mean field states at all.
Moreover, the number of zeroes of those kinematical factors in the
HF determinant as a function of any one of the identical particle
coordinates, are just $(q-1)N_{e}$ were $N_{e}$ is the number of
particles. \ This property means that the mentioned number of
zeroes is identical to the one which which $q$-power defines the
variational Laughlin states in the Haldane-Rezayi scheme. \
Therefore, substituting the crystalline like factors in the mean
field determinantal state by the liquid like ones appearing in the
Laughlin wave function, leads to the same short range behavior (a
zero of order $q$) when any two particles join one another.
However, in addition the presence of the determinantal functions
remains representing the crystalline information associated with
the optimization of the mean field problem. Therefore,  the
proposed state have the chance of slightly improving the energies
per particle of the Laughlin ones. \ This possibility opens a way
of checking whether or not the long standing claims about the
breaking of the translation invariance in QHE ground state can
have a confirmation which in addition retains the demonstrated
robustness of the Laughlin or Jain descriptions.

The correctness of such results can support the relevance  of the
mean field solutions, long time ago proposed by one of the present
authors (F.C.),\ \ as correct precursor states within a
perturbative description of the \ FQHE ground states. In it,  the
inclusion of correlations in each order could approach the results
in the real  FQHE ground states. \ \ Evaluations of the energy per
particle in the examined states are planned to be considered.

In the first Section, the single particle orbitals solving the HF
problem for the two considered classes of states are explicitly
expressed in the Haldane-Rezayi scheme for the periodic reduction
of the $HF$ problem. These states were derived in Ref.
\onlinecite{ccpm} and the notation and definitions here closely
follows the ones in that work. Thus, they should be considered as
complementary articles. \ The next section aboard the main subject
of the paper, the construction of the variational states, which
are motivated by the special structure of the one mean field one
particle orbitals. \ \ \ \

\section{The HF single particle states at $\nu=1/q$ in the Haldane-Rezayi scheme}

Let us consider in this first section the determination of the analytic form
of the one particle mean field orbitals for filling factors $\nu=1/q.$
\ Expression for these orbitals were obtained in the previous paper
[\onlinecite{ccpm}]. \ The new representation will be more appropriate (but
not essentially needed) for the discussion below, in which we want to consider
in periodic boundary conditions.

A formula given in the periodic boundary conditions scheme \ developed by
Haldane and Rezayi in Ref. [\onlinecite{halrez}], will be helpful. It
expresses any wave-function in the fist Landau level which satisfy periodic
boundary conditions, in terms of its zeros laying inside the first region of
periodicity. The expression is
\[
\varphi_{L}(\mathbf{x})=\exp(\widetilde{k\ }z^{\ast}-\frac{y^{2}}{2r_{o}^{2}%
})\prod_{\varsigma=1}^{N_{o}}\vartheta_{1}(\frac{\pi}{L_{1}}(z^{\ast
}-z_{\varsigma})|-\tau^{\ast}).
\]
The constant $\widetilde{k\ }$ is purely imaginary, and for large $A$ as in
our case is almost continuous. The $z_{\varsigma}$ are the complex coordinates
of the zeros of the wave-functions, which should be a total number equal to
the number of flux quanta $N_{o}$ piercing the area $A.$ The periodic
conditions are imposed on the boundary of this zone (that we will call first
periodicity zone). In our situation, having a sample of area $A$ with large
sides of length $L_{1}$ and $L_{2}$ and forming between them an angle of
$\frac{2\pi}{6}$, the parameter $\tau^{\ast}$ has the form
\[
\tau^{\ast}=\frac{L_{1}}{L_{2}}\exp(-\frac{2\pi i}{6}).
\]
We will, also assume that $L_{1}=L_{2}.$ In the present case the complex
conjugate variable $\ z^{\ast}$ appears in place of $z$ because the difference
in conventions with reference [\onlinecite{halrez}]. Also, $\varphi
_{L}(\mathbf{x})$ is written in the Landau gauge. After performing the
necessary gauge transformation, the corresponding functions in the axial gauge
take the forms
\[
\varphi(\mathbf{x})=\exp(k\ z^{\ast}+\frac{z^{\ast2}}{4r_{o}^{2}}%
-\frac{\mathbf{x}^{2}}{4r_{o}^{2}})\prod_{\varsigma=1}^{N_{o}}\vartheta
_{1}(\frac{\pi}{L_{1}}(z^{\ast}-z_{\varsigma})|-\tau^{\ast}).
\]

Let us consider below the determination of the functions for each of the two
cases of particle number per unit cell $\gamma=1$ and $\gamma=1/2.$

\subsection{Case $\gamma=1$}

First we fix $(q-1)$ zeros to each of the points of the lattice of periodicity
of the density $\mathbf{R}$ being inside the first periodicity zone. Also one
zero will be chosen for each cell but at positions differing from each point
$\mathbf{R}$ by a constant $C_{\mathbf{k}}$ to be determined. The first
condition assures that the wave functions have zero of order $(q-1)$ at any
point of the lattice $\mathbf{R.}$\textbf{\ }. Then, the functions have the
form
\begin{align*}
\varphi_{\mathbf{k}}^{(0)}(\mathbf{x})  &  =\exp(k\ z^{\ast}+\frac{z^{\ast2}%
}{4r_{o}^{2}}-\frac{\mathbf{x}^{2}}{4r_{o}^{2}})\times\\
&  \prod_{R}\left\{  \left(  \vartheta_{1}(\frac{\pi}{L_{1}}(z^{\ast}-R^{\ast
})|-\tau^{\ast})\right)  ^{(q-1)}\vartheta_{1}(\frac{\pi}{L_{1}}(z^{\ast
}-R^{\ast}-C_{\mathbf{k}})|-\tau^{\ast})\right\}  ,
\end{align*}
where according with the conventions employed the complex coordinates for the
vectors of the lattice $\mathbf{R}$ are defined by the integers $n_{1}$ and
$n_{2}\ $ as
\[
R=n_{1}a+n_{2}\ a_{2},\;\;\;\;R^{\ast}=n_{1}a+n_{2}\ a_{2}^{\ast},
\]
in which $a$ is the size of the unit cell side defined in Ref. [\onlinecite
{ccpm}], and for any vector $\mathbf{d}=(d_{1},d_{2},0)$ we write
\begin{equation}
d=d_{1}+d_{2}i,\ \;\;d^{\ast}=d_{1}-d_{2}i.
\end{equation}

The number of cells in the first periodicity region and their sizes are
related by .
\begin{align*}
L_{1}  &  =a\text{ }N_{1},\;L_{2}=a\text{ }N_{2},\\
N_{1}  &  =N_{2},\;\;\;N_{1}=\text{even\ number},
\end{align*}
and the allowed values of $n_{1}$ and $n_{2}$ in the first periodicity region
are
\begin{align*}
n_{1}  &  \in\{-\frac{N_{1}}2,-\frac{N_{1}}2+1,..,0,...,\frac{N_{1}}2-1\},\\
n_{2}  &  \in\{-\frac{N_{2}}2,-\frac{N_{2}}2+1,..,0,...,\frac{N_{2}}2-1\}.
\end{align*}

Lets us now impose the condition on the functions to pertain to a given
$q$-multiplet which is characterized by the eigenvalues $exp(-i\mathbf{k.R})$
of the translation operators $T_{\mathbf{R}}$ in the lattice vectors
$\mathbf{R.}$ For this purpose it is sufficient to consider the translations
under the unit cell vectors $\mathbf{a}_{1}$ and $\mathbf{a}_{2}$. The action
of a general translation operator over a function of the first Landau level
having an analytic part defined in Ref. [\onlinecite{ccpm}] can be written as
\[
T_{\mathbf{R}}F(z^{*})\exp(-\frac{\mathbf{x}^{2}}{4r_{o}^{2}})=\exp
(-\frac{RR^{*}}{4r_{o}^{2}}+\frac{R\ z^{*}}{2r_{o}^{2}})F(z^{*}-R^{*}%
)\exp(-\frac{\mathbf{x}^{2}}{4r_{o}^{2}}).
\]

Employing this relation in performing a translation in $\mathbf{a}_{1}$ we
have
\begin{align*}
T_{\mathbf{a}_{1}}\varphi_{\mathbf{k}}^{(0)}(\mathbf{x})  &  =\exp
(\widetilde{k}\ z^{*}+\frac{z^{*2}}{4r_{o}^{2}}-\frac{\mathbf{x}^{2}}%
{4r_{o}^{2}})\times\exp(-\widetilde{k}\ a)\\
&  \prod_{R}\left\{  \left(  \vartheta_{1}(\frac\pi{L_{1}}(z^{*}%
-R^{*}-a)|-\tau^{*})\right)  ^{(q-1)}\vartheta_{1}(\frac\pi{L_{1}}(z^{*}%
-R^{*}-a-C_{\mathbf{k}})\right\}  |-\tau^{*}).
\end{align*}

But, under the shift $R^{*}\rightarrow R^{*}+a$, almost all the appearing
$\vartheta_{1}$ functions transform one in another. Only the ones having
$n_{1}=(\frac{N_{1}}2-1)$ for all values of $n_{2}$, transform in different
ones not appearing and $n_{1}=\frac{N_{1}}2.$ But, representing $\frac{N_{1}%
}2$ as $-\frac{N_{1}}2+N_{1}$ and using the translation property of the
elliptic function
\begin{equation}
\vartheta_{1}(w^{*}-\pi|-\tau^{*})=-\vartheta_{1}(w^{*}|-\tau^{*}),
\label{per1}%
\end{equation}
allows to write
\[
T_{\mathbf{a}_{1}}\varphi_{\mathbf{k}}^{(0)}(\mathbf{x})=(-1)^{qN_{2}}%
\exp(-\widetilde{k}\ a)\varphi_{\mathbf{k}}^{(0)}(\mathbf{x}).
\]
Remembering that $N_{2}$ was selected as an even number we have $(-1)^{qN_{2}%
}=1$. Furthermore, after choosing the imaginary parameter $\widetilde{k}$ as
$\widetilde{k}=\mathbf{k}.\mathbf{a}_{1}i\ $ it follows
\[
T_{\mathbf{a}_{1}}\varphi_{\mathbf{k}}^{(0)}(\mathbf{x})=\exp(-\mathbf{k}%
.\mathbf{a}_{1}i)\;\varphi_{\mathbf{k}}^{(0)}(\mathbf{x}),
\]
which is one of the eigenvalue conditions obeyed by the $q$ multiplets. Next,
it rests to impose on $\varphi_{\mathbf{k}}^{(0)}$ the character of being an
eigen-function of $T_{\mathbf{a}_{2}}.$ Performing the translation in
$\mathbf{a}_{2}$follows
\begin{align}
T_{\mathbf{a}_{2}}\varphi_{\mathbf{k}}^{(0)}(\mathbf{x})  &  =\exp
(\widetilde{k}\ z^{*}+\frac{z^{*2}}{4r_{o}^{2}}-\frac{\mathbf{x}^{2}}%
{4r_{o}^{2}})\times\exp(\frac{a_{2}^{2*}-a^{2}}{4r_{o}^{2}}+\ \frac{z^{*}%
}{2r_{o}^{2}}(a_{2}-a_{2}^{*})-\widetilde{k}a_{2}^{*})\times\label{Ta2}\\
&  \prod_{R}\left\{  \left(  \vartheta_{1}(\frac\pi{L_{1}}(z^{*}-R^{*}%
-a_{2}^{*})|-\tau^{*})\right)  ^{(q-1)}\vartheta_{1}(\frac\pi{L_{1}}%
(z^{*}-R^{*}-a_{2}^{*}-C_{\mathbf{k}})\right\}  |-\tau^{*}),\nonumber
\end{align}
where as before, almost all the $\vartheta_{1}$ transform among themselves and
solely the ones having $n_{2}$=$(\frac{N_{2}}2-1)$ pass to have $n_{2}$%
=$\frac{N_{2}}2.$ Thus, again representing $\frac{N_{2}}2=-\frac{N_{2}}%
2+N_{2}$ and making use the other translation property of the elliptic
function $\vartheta_{1}$
\begin{equation}
\vartheta_{1}(w^{*}-\pi\tau^{*}|-\tau^{*})=-\exp(-i(2w^{*}-\pi\tau
^{*}))\vartheta_{1}(w^{*}|-\tau^{*}) \label{per2}%
\end{equation}
the relation (\ref{Ta2}) takes the form
\begin{align*}
T_{\mathbf{a}_{2}}\varphi_{\mathbf{k}}^{(0)}(\mathbf{x})  &  =(-1)^{qN_{1}%
}\exp(D_{\mathbf{k}})\;\varphi_{\mathbf{k}}^{(0)}(\mathbf{x})\\
D_{\mathbf{k}}  &  =-\frac{2\pi i}aC_{\mathbf{k}}-i\mathbf{k}.\mathbf{a}%
_{1}\tau^{*}-\frac{a^{2}}{4r_{o}^{2}}(1-\tau^{2*})-i\pi q,
\end{align*}
which after imposing the condition on the function of being an eigen-function
of $T_{\mathbf{a}_{2}}$ with eigenvalue $\exp(-i\mathbf{k}.\mathbf{a}_{1})$,
that is $D_{\mathbf{k}}=-i\mathbf{k}.\mathbf{a}_{1},$ the constant
$C_{\mathbf{k}}$ is fixed to be
\[
C_{\mathbf{k}}=\frac a\pi(\mathbf{k}.\mathbf{a}_{2}-\mathbf{k}.\mathbf{a}%
_{1}\tau^{*})+\frac{ia\ q}{2\sqrt{3}}(1-\tau^{2*})-\frac{q\ a}2.
\]
which assures the resting eigenvalue condition
\[
T_{\mathbf{a}_{2}}\varphi_{\mathbf{k}}^{(0)}(\mathbf{x})=\exp(-i\mathbf{k}%
.\mathbf{a}_{1})\varphi_{\mathbf{k}}^{(0)}(\mathbf{x}).
\]

Finally, let us consider that the function inside each $q$ multiplet showing a
zero of order $(q-1)$ should be unique. However, each multiplet is also
uniquely determined by the value of the quasi-momentum $\mathbf{k}$ defining
its eigenvalues $\exp(-i\mathbf{k}.\mathbf{a}_{1})$ and $\exp(-i\mathbf{k}%
.\mathbf{a}_{2})$ under the translations $T_{\mathbf{a}_{1}}$ and
$T_{\mathbf{a}_{2}},$ respectively. Therefore, the just constructed function
$\varphi_{\mathbf{k}}^{(0)}(\mathbf{x})$ should be unique one showing zeros of
order $(q-1)$ laying in the multiplet indexed by $\mathbf{k.}$

\subsection{Case $\gamma=1/2$}

Since the case $\gamma=1/2,$ has some additional subtleties, let us also
consider it in detail. In this case we fix $\frac{(q-1)}{2}$ zeros to each
point of the lattice $\mathbf{R}$ and again shift the zero of the other factor
in the constant to be determined $C_{(\mathbf{k,\sigma)}}.$ Then,
\begin{align*}
\varphi_{(\mathbf{k,\sigma)}}^{(0)}(\mathbf{x})  &  =\exp(k\ z^{\ast}%
+\frac{z^{\ast2}}{4r_{o}^{2}}-\frac{\mathbf{x}^{2}}{4r_{o}^{2}})\times\\
&  \prod_{R}\left\{  \left(  \vartheta_{1}(\frac{\pi}{L_{1}}(z^{\ast}-R^{\ast
})|-\tau^{\ast})\right)  ^{\frac{(q-1)}{2}}\right\}  \prod_{R_{\sigma}%
}\left\{  \vartheta_{1}(\frac{\pi}{L_{1}}(z^{\ast}-R_{\sigma}^{\ast
}-C_{(\mathbf{k,\sigma)}})|-\tau^{\ast})\right\}  .
\end{align*}

The position of the zero of order $\frac{(q-1)}{2}$ are at all the points
$R^{\ast}=n_{1}a+n_{2}\ a_{2}^{\ast}$, being inside the first periodicity
zone. However, in order to be able of generating the doublet of functions for
the two values of $\sigma,$ the arguments of the functions not showing zeros
exactly at the lattice points $R$ will be defined by the two set of lattice
vectors
\begin{align*}
R_{\sigma}  &  =n_{1}^{\sigma}a+n_{2}\ a_{2},R^{\ast}=[2m_{1}+\frac
{(1-\sigma)}{2})]a+n_{2}\ a_{2}^{\ast},\quad\\
m_{1}  &  =0,\pm1,\pm2,...,\;\sigma=\pm1.
\end{align*}
where $a$ \ is the size of the unit cell defined in Ref. [\onlinecite{ccpm}]
for $\phi/\phi_{o}=q/2.$ \ The union of the two lattices gives the l$\arg$er
lattice $R$.

In this case the sizes of the first periodicity zone and the number of cells
per side satisfy
\begin{align*}
L_{1}  &  =a\text{ }N_{1},L_{2}=a\text{ }N_{2},\\
N_{1}  &  =N_{2}\\
N_{1}  &  =4t+2,t=1,2,...,
\end{align*}
and now the range of allowed values for the indices defining the lattices
$\mathbf{R}_{\sigma}$ are
\begin{align*}
n_{1}^{+}  &  \in\{-(\frac{N_{1}}2-1),-(\frac{N_{1}}2-1)+2,..,0,...,\frac
{N_{1}}2-3,\frac{N_{1}}2-1\},\\
n_{1}^{-}  &  \in\{-\frac{N_{1}}2,-\frac{N_{1}}2+2,..,0,...,\frac{N_{1}%
}2-2\},\\
n_{2}  &  \in\{-\frac{N_{2}}2,-\frac{N_{2}}2+1,..,0,...,\frac{N_{2}}2-1\}.
\end{align*}

Lets us now impose the condition that the functions pertains to the space
formed by union of the two $q$-multiplets associated to $\sigma=\pm1.$ This
set is fixed by the momenta $\mathbf{k}$ in Brillouin cell of the reciprocal
lattice associated to the lattice vectors $2\mathbf{R.}$ The action of
$T_{\mathbf{a}_{1}}$ can be written as
\begin{align*}
T_{\mathbf{a}_{1}}\varphi_{(\mathbf{k,}\sigma\mathbf{)}}^{(0)}(\mathbf{x})  &
=\exp(\widetilde{k}\ z^{*}+\frac{z^{*2}}{4r_{o}^{2}}-\frac{\mathbf{x}^{2}%
}{4r_{o}^{2}})\times\exp(-\widetilde{k}\ a)\times\\
&  \prod_{R}\left\{  \left(  \vartheta_{1}(\frac\pi{L_{1}}(z^{*}%
-[(n_{1}+1)a+n_{2}a_{2}^{*}])|-\tau^{*})\right)  ^{\frac{(q-1)}2}\right\}
\times\\
&  \prod_{R_{\sigma}}\left\{  \vartheta_{1}(\frac\pi{L_{1}}(z^{*}%
-[(n_{1}^{\sigma}+1)a+n_{2}a_{2}^{*}])-C_{\mathbf{k}})|-\tau^{*})\right\}  .
\end{align*}

Note, that for $\sigma=-1$ all the lattice $\mathbf{R}_{-1}$ transforms in the
lattice $\mathbf{R}_{+1}.$ For the factor showing the zeros exactly at the
lattice $\mathbf{R,}$ only the $\vartheta_{1}$ functions having $n_{1}%
=(\frac{N_{1}}{2}-1)$ for all values of $n_{2}$ do not transform in other
functions appearing in the product. But, representing for them $\frac{N_{1}%
}{2}$ as $-\frac{N_{1}}{2}+N_{1}$ and employing the symmetry property
(\ref{per1}), it is possible to write
\[
T_{\mathbf{a}_{1}}\varphi_{(\mathbf{k,\sigma)}}^{(0)}(\mathbf{x}%
)=\exp(-\widetilde{k}\ a)\varphi_{(\mathbf{k,-\sigma)}}^{(0)}(\mathbf{x}),
\]
a relation which indicates that the translation in\textbf{\ }$\mathbf{a}_{1}$
interchanges the values of $\sigma$. Fixing the imaginary parameter
$\widetilde{k}$ as $\widetilde{k}=\mathbf{k}.\mathbf{a}_{1}i\ $ allows to fix
the transformation property
\[
T_{\mathbf{a}_{1}}\varphi_{(\mathbf{k,\sigma)}}^{(0)}(\mathbf{x}%
)=\exp(-\mathbf{k}.\mathbf{a}_{1}i)\varphi_{(\mathbf{k,-\sigma)}}%
^{(0)}(\mathbf{x}).
\]
Thus , the functions for the two values of $\sigma$ transform among
themselves. This property, in turns implies that the functions are
eigen-functions of the lattice formed by the vectors $2\mathbf{R}$. Let us
inspect finally the action on $\varphi_{\mathbf{k}}^{(0)}$ of the operator
$T_{\mathbf{a}_{2}}.$%
\begin{align*}
T_{\mathbf{a}_{2}}\varphi_{(\mathbf{k,\sigma)}}^{(0)}(\mathbf{x})  &
=\exp(\widetilde{k}\ (z^{\ast}-a_{2}^{\ast})+\frac{(z^{\ast}-a_{2}^{\ast}%
)^{2}}{4r_{o}^{2}}-\frac{\mathbf{x}^{2}}{4r_{o}^{2}})\times\\
&  \exp(-\frac{a_{2}a_{2}^{\ast}}{4r_{o}^{2}})+\frac{a_{2}z^{\ast}}{2r_{o}%
^{2}})\times\\
&  \prod_{R}\left\{  \left(  \vartheta_{1}(\frac{\pi}{L_{1}}(z^{\ast}%
-[n_{1}a+(n_{2}+1)a_{2}^{\ast}])|-\tau^{\ast})\right)  ^{\frac{(q-1)}{2}%
}\right\}  \times\\
&  \prod_{R_{\sigma}}\left\{  \vartheta_{1}(\frac{\pi}{L_{1}}(z^{\ast}%
-[n_{1}^{\sigma}a+(n_{2}+1)a_{2}^{\ast}])-C_{\mathbf{k}})|-\tau^{\ast
})\right\}  .
\end{align*}
In this case, only the functions for the indices $n_{2}$=$(\frac{N_{2}}{2}-1)$
for any value of $n_{1}$transform in other $\vartheta_{1}$ terms being absent
in the original product. After again representing $\frac{N_{2}}{2}%
=-\frac{N_{2}}{2}+N_{2}$ and using (\ref{per2}) it is possible to impose the
eigenvalue relation
\[
T_{\mathbf{a}_{2}}\varphi_{(\mathbf{k,\sigma)}}^{(0)}(\mathbf{x}%
)=\exp(-i\mathbf{k}.\mathbf{a}_{2})\varphi_{(\mathbf{k,\sigma)}}%
^{(0)}(\mathbf{x}),
\]
after selecting the constant $C_{(\mathbf{k,\sigma)}}$ as given by
\[
C_{(\mathbf{k,}\sigma\mathbf{)}}=\frac{a}{\pi}(\mathbf{k}.\mathbf{a}%
_{2}-\mathbf{k}.\mathbf{a}_{1}\tau^{\ast})+\frac{ia\ q}{2\sqrt{3}}%
(1-\tau^{2\ast})-\frac{(q-\sigma)\ a}{2}+a\,q\,\alpha
\]
where $\alpha$ is equal to $0$ or $1$ if $\frac{N_{1}}{2}$ is even or odd respectively.

In this case it happens that the function inside each $q$ multiplet
constructed in Ref. \onlinecite{ccpm} and showing a zero of order
$\frac{(q-1)}{2}$ should be unique for each of the two values of $\sigma$ if
the set of associated equations fix them. Therefore, there is a doublet of
functions satisfying the vanishing conditions. Then, the two functions
$\varphi_{(\mathbf{k,+)}}^{(0)}(\mathbf{x})$ and $\varphi_{(\mathbf{k,-)}%
}^{(0)}(\mathbf{x})$ determined here, uniquely expands this doublet for a
given value of $\mathbf{k.}$ However, which linear combination defining the
specific functions fixed in each of the multiplets formed by the functions
$\chi_{\mathbf{k}}^{(r,\sigma)}$ defined in Ref. [\onlinecite{ccpm}] could be
easily determined.

\section{Trial states including short range as well as crystalline correlations}

\subsection{Case $\gamma=1/2$\ }

Changing the order with respect to the previous section, let us consider first
the Slater determinant related with the $HF$ problem in the case $\gamma
=\frac{1}{2}$ \ for which the HF single particle orbitals are $\varphi
_{(\mathbf{k,\sigma)}}^{(0)}(\mathbf{x})$. \ The mean field determinantal
state in this case has the form
\[
\Psi_{hf}^{(\frac{1}{2})}=\Psi_{hf}[z_{1}^{\ast},z_{2}^{\ast},...z_{N_{e}%
}^{\ast},z_{1},z_{2},...z_{N_{e}}]=Det\text{ }[\varphi_{(\mathbf{k}%
_{i}\mathbf{,\sigma}_{i}\mathbf{)}}^{(0)}(\mathbf{x}_{j})],
\]
where the wavevectors $\mathbf{k}_{i}$ pertain to the two degenerate bands
(for both values of $\sigma$), defining the ground state of the $HF$ problem.
These momenta satisfy the periodicity condition of the Haldane-Rezayi
procedure and their number is equal to one half of the number of electrons in
the problem. \ As shown in the previous section the orbitals have the
structure%
\begin{align*}
\varphi_{(\mathbf{k,\sigma)}}^{(0)}(\mathbf{x}) &  =\prod_{R}\left\{  \left(
\vartheta_{1}(\frac{\pi}{L_{1}}(z^{\ast}-R^{\ast})|-\tau^{\ast})\right)
^{\frac{(q-1)}{2}}\right\}  \chi_{(\mathbf{k,\sigma)}}^{(0)}(z^{\ast}%
)\exp(\frac{z^{\ast2}}{4r_{o}^{2}}-\frac{\mathbf{x}^{2}}{4r_{o}^{2}})\\
&  =P(z^{\ast})\text{ }\chi_{(\mathbf{k,\sigma)}}^{(0)}(z^{\ast}%
)\exp(k\ z^{\ast}+\frac{z^{\ast2}}{4r_{o}^{2}}-\frac{\mathbf{x}^{2}}%
{4r_{o}^{2}})\\
P(z^{\ast}) &  =\prod_{R}\left\{  \left(  \vartheta_{1}(\frac{\pi}{L_{1}%
}(z^{\ast}-R^{\ast})|-\tau^{\ast})\right)  ^{\frac{(q-1)}{2}}\right\}  \\
\chi_{(\mathbf{k,\sigma)}}^{(0)}(z^{\ast}) &  \text{=}\exp(k\ z^{\ast}%
)\prod_{R_{\sigma}}\left\{  \vartheta_{1}(\frac{\pi}{L_{1}}(z^{\ast}%
-R_{\sigma}^{\ast}-C_{(\mathbf{k,\sigma)}})|-\tau^{\ast})\right\}  .
\end{align*}
where all \ the dependences of the momentum quantum numbers and the index
$\ \sigma$ $\ $ are contained in the new function $\chi.$

The product of $\vartheta_{1}$ functions: \ $P(z^{\ast})$\ \ is an analytical
function of $\ z^{\ast}$ showing zeroes at all the points of the lattice
$\cite{ccpm}$. \ Its number of zeroes as a function of $\ z^{\ast}$ is given
by $\frac{(q-1)}{2},$ \ times the number of points of the lattice $R$.
\ \ But, this last value \ is equal to twice the number of particles of the
system $N_{e}$ (racall that each cell has $q/2$ flux quantain this
$\gamma=1/2$ case).   \ The just described structure of the single particle
$HF$ orbitals leads to the following form for the Slater determinant%

\begin{align}
\Psi_{hf}^{(\frac{1}{2})}  &  =Det\text{ }[\varphi_{(\mathbf{k}_{i}%
\mathbf{,\sigma}_{i}\mathbf{)}}^{(0)}(\mathbf{x}_{j})]\nonumber\\
&  =(\prod_{k=1,2,..N_{e}}P(z_{k}^{\ast}))Det\text{ }[\chi_{(\mathbf{k}%
_{i}\mathbf{,\sigma}_{i}\mathbf{)}}^{(0)}(z_{k}^{\ast})]\exp(\sum
_{i=1,2...N_{e}}(k\ z_{i}^{\ast}+\frac{z_{i}^{\ast2}}{4r_{o}^{2}}%
-\frac{\mathbf{x}_{i}^{2}}{4r_{o}^{2}}))\nonumber\\
&  =\Phi(z_{1}^{\ast},z_{2}^{\ast},...z_{N_{e}}^{\ast})D[z_{1}^{\ast}%
,z_{2}^{\ast},...z_{N_{e}}^{\ast}]\exp(\sum_{i=1,2...N_{e}}(\frac{z_{i}%
^{\ast2}}{4r_{o}^{2}}-\frac{\mathbf{x}_{i}^{2}}{4r_{o}^{2}}))\label{phi}\\
\Phi(z_{1}^{\ast},z_{2}^{\ast},...z_{N_{e}}^{\ast})  &  =\prod_{R}\left\{
\left(  \vartheta_{1}(\frac{\pi}{L_{1}}(z^{\ast}-R^{\ast})|-\tau^{\ast
})\right)  ^{\frac{q-1}{2}}\right\} \nonumber\\
D[z_{1}^{\ast},z_{2}^{\ast},...z_{N_{e}}^{\ast}]  &  =\exp(\sum_{i=1,2...N_{e}%
}k\ z_{i}^{\ast})Det\text{ }[\chi_{(\mathbf{k}_{i}\mathbf{,\sigma}%
_{i}\mathbf{)}}^{(0)}(z_{k}^{\ast})].\nonumber
\end{align}
\ \

It can be noticed that the functions $\Phi,$ \thinspace\ when considered as
depending of any of the electron coordinates $z^{\ast},$ show a number of
zeroes equal to $(q-1)N_{e}$. \ Since the determinant $Det$ $[\chi
_{(\mathbf{k}_{i}\mathbf{,\sigma}_{i}\mathbf{)}}^{(0)}(z_{j}^{\ast})]$ in
addition is showing  a number \ $N_{e}$ \ zeroes, the total number of \ zeroes
of the function $\Psi_{hf}^{(\frac{1}{2})}$ with respect to any variable is
equal to the number of flux quanta $qN_{e}$ passing through the periodicity
area . \ Therefore, this basic requirement of the periodic representation of
the space of single particle wave functions is satisfied. \ \

Thus, the above  property \ directly suggests the main idea addressed in this
paper: to propose variational states which are close connected with the states
being the counterpart of the homogeneous Laughlin \  in the Haldane-Rezayi
scheme. After transforming to the axial gauge the form of these periodic
Laughlin states (which were determined in the Landau gauge in Ref \ [%
$\backslash$%
onlinecite\{halrez\}])  is
\begin{align}
\Psi_{L} &  =\Psi_{L}[z_{1}^{\ast},z_{2}^{\ast},...z_{N_{e}}^{\ast}%
,z_{1},z_{2},...z_{N_{e}}]=\exp(\sum_{i=1,2...N_{e}}(\frac{z_{i}^{\ast2}%
}{4r_{o}^{2}}-\frac{\mathbf{x}_{i}^{2}}{4r_{o}^{2}}))\times\prod
_{\substack{i<j\\i,j=1,2...N_{e}}}\left\{  \vartheta_{1}(\frac{\pi}{L_{1}%
}(z_{i}^{\ast}-z_{j}^{\ast})|-\tau^{\ast})\right\}  ^{q}\nonumber\\
&  =\Phi_{L}(z_{1},z_{2},...z_{N_{e}})D_{L}[z_{1},z_{2},...z_{N_{e}}]\exp
(\sum_{i=1,2...N_{e}}(\frac{z_{i}^{\ast2}}{4r_{o}^{2}}-\frac{\mathbf{x}%
_{i}^{2}}{4r_{o}^{2}}))\nonumber\\
\Phi_{L}(z_{1}^{\ast},z_{2}^{\ast},...z_{N_{e}}^{\ast}) &  =\left\{
\vartheta_{1}(\frac{\pi}{L_{1}}(Z^{\ast}-\overline{R}^{\ast})|-\tau^{\ast
})\right\}  ^{q-1}\prod_{\substack{i<j\\i,j=1,2...N_{e}}}\left\{
\vartheta_{1}(\frac{\pi}{L_{1}}(z_{i}^{\ast}-z_{j}^{\ast})|-\tau^{\ast
})\right\}  ^{q-1}=D_{L}^{q-1}(z_{1}^{\ast},z_{2}^{\ast},...z_{N_{e}}^{\ast
})\label{phil}\\
D_{L}[z_{1}^{\ast},z_{2}^{\ast},...z_{N_{e}}^{\ast}] &  =\vartheta_{1}%
(\frac{\pi}{L_{1}}(Z^{\ast}-\overline{R}^{\ast})|-\tau^{\ast})\prod
_{\substack{i<j\\i,j=1,2...N_{e}}}\vartheta_{1}(\frac{\pi}{L_{1}}(z_{i}^{\ast
}-z_{j}^{\ast})|-\tau^{\ast})\nonumber\\
Z &  =\sum_{i=1,,2,..N_{e}}z_{i},\text{ \ \ }\overline{R}=\sum_{R}R\nonumber
\end{align}
where the center of mass wave function \ defined in [\onlinecite{halrez}] has
been chosen to have its $q$ zeroes at the point $Z=\overline{R},$ That is, at
the point given by the sum of all the lattice $R$ coordinates. \ The center of
mass momenta, also defined in [\onlinecite{halrez}] is also taken to vanish.
The above representation, indicates that  the \ factors $\Phi$ in
(\ref{phi}),  and $\Phi_{L}$ in (\ref{phil}), when considered as functions of
the coordinates $z_{i}^{\ast}$ of any particular electron, have the same
number of zeroes equal to $(q-1)N_{e}.$ The same equality in the number of
zeroes for any particular coordinate $z_{i}^{\ast}$ also show  the functions
$D$ and $D_{L}.$ \ Therefore, the noticed identity in the number of zeroes of
the functions $\Phi$ and $\Phi_{L}$, directly leads to the idea of
substituting the function $\Phi$ \ in \ \ref{phi} \ by the function $\Phi_{L}$
, in order to construct new variational states. These new wave functions will
show the same order of the zeroes when any two particle equalize their
coordinates, but in addition they are expected to  also retain the sort
crystalline correlations which optimized the mean field problem.

Considering the above remarks, the new  variational states are proposed in the
form
\begin{align*}
\Psi_{N}^{(\frac{1}{2})} &  =\Psi_{N}[z_{1}^{\ast},z_{2}^{\ast},...z_{N_{e}%
}^{\ast},z_{1},z_{2},...z_{N_{e}}]=\Phi_{L}(z_{1}^{\ast},z_{2}^{\ast
},...z_{N_{e}}^{\ast})D^{(\frac{1}{2})}[z_{1}^{\ast},z_{2}^{\ast},...z_{N_{e}%
}^{\ast}]\exp(\sum_{i=1,2...N_{e}}(\frac{z_{i}^{\ast2}}{4r_{o}^{2}}%
-\frac{\mathbf{x}_{i}^{2}}{4r_{o}^{2}})),\\
\Phi_{L}(z_{1}^{\ast},z_{2}^{\ast},...z_{N_{e}}^{\ast}) &  =\left\{
\vartheta_{1}(\frac{\pi}{L_{1}}(Z^{\ast}-\overline{R}^{\ast})|-\tau^{\ast
})\right\}  ^{q-1}\prod_{\substack{i<j\\i,j=1,2...N_{e}}}\left\{
\vartheta_{1}(\frac{\pi}{L_{1}}(z_{i}^{\ast}-z_{j}^{\ast})|-\tau^{\ast
})\right\}  ^{q-1},\\
D^{(\frac{1}{2})}(z_{1}^{\ast},z_{2}^{\ast},...z_{N_{e}}^{\ast}) &  =\exp
(\sum_{i=1,2...N_{e}}k\ z_{i}^{\ast})Det\text{ }[\chi_{(\mathbf{k}%
_{i}\mathbf{,\sigma}_{i}\mathbf{)}}^{(0)}(z_{k}^{\ast})].
\end{align*}

\ \ Let us now argue that this many particle variational state when considered
as a function of any of its coordinates

satisfies the periodicity condition of the problem.For this purpose it is
useful tow write the HF determinantal estate in the form%

\begin{align*}
\Psi_{N}^{(\frac{1}{2})} &  =\frac{\Phi_{L}(z_{1}^{\ast},z_{2}^{\ast
},...z_{N_{e}}^{\ast})}{\Phi(z_{1}^{\ast},z_{2}^{\ast},...z_{N_{e}}^{\ast}%
)}\Psi_{hf}^{(\frac{1}{2})}[z_{1}^{\ast},z_{2}^{\ast},...z_{N_{e}}^{\ast
},z_{1}^{\ast},z_{2}^{\ast},...z_{N_{e}}^{\ast}],\\
&  =\left\{  \frac{F_{L}(z_{1}^{\ast},z_{2}^{\ast},...z_{N_{e}}^{\ast}%
)}{F(z_{1}^{\ast},z_{2}^{\ast},...z_{N_{e}}^{\ast})}\right\}  ^{q-1}\Psi
_{hf}^{(\frac{1}{2})}[z_{1}^{\ast},z_{2}^{\ast},...z_{N_{e}}^{\ast}%
,z_{1}^{\ast},z_{2}^{\ast},...z_{N_{e}}^{\ast}],\\
&  =\left\{  \eta(z_{1}^{\ast},z_{2}^{\ast},...z_{N_{e}}^{\ast})\right\}
^{q-1}\Psi_{hf}^{(\frac{1}{2})}[z_{1}^{\ast},z_{2}^{\ast},...z_{N_{e}}^{\ast
},z_{1}^{\ast},z_{2}^{\ast},...z_{N_{e}}^{\ast}],\\
F(z_{1}^{\ast},z_{2}^{\ast},...z_{N_{e}}^{\ast}) &  =\prod_{j}\prod
_{R}\left\{  \left(  \vartheta_{1}(\frac{\pi}{L_{1}}(z_{j}^{\ast}-R^{\ast
})|-\tau^{\ast})\right)  ^{\frac{q-1}{2}}\right\}  ,\\
F_{L}(z_{1}^{\ast},z_{2}^{\ast},...z_{N_{e}}^{\ast}) &  =\left\{
\vartheta_{1}(\frac{\pi}{L_{1}}(Z^{\ast}-\overline{R}^{\ast})|-\tau^{\ast
})\right\}  ^{q-1}\prod_{\substack{i<j\\i,j=1,2...N_{e}}}\left\{
\vartheta_{1}(\frac{\pi}{L_{1}}(z_{i}^{\ast}-z_{j}^{\ast})|-\tau^{\ast
})\right\}  .
\end{align*}
But, the form of the magnetic translations in the axial gauge (as defined by
its action on the analytic factors $A$ of the exponential $\exp(-\frac
{\mathbf{x}^{2}}{4r_{o}^{2}})$) is%
\[
T_{R}A(z^{\ast})=\exp(-\frac{RR^{\ast}}{4r_{o}^{2}}+\frac{Rz^{\ast}}%
{2r_{o}^{2}})A(z^{\ast}-R^{\ast}).
\]
and  $\Psi_{hf}$ by construction, is invariant under the magnetic
translations\ of $\ $any\ of $z_{i}^{\ast}$ variables\ in the $L_{1}$ and
$L_{1}\tau^{\ast},$ corresponding to the elementary shifts in the lattice in
which periodicity conditions are imposed \ Therefore, the variation of the
wave function $\Psi_{N}^{(\frac{1}{2})}$ will be determined by the
corresponding change in the analytic factor $\eta,$ under the shifts
$\eta(z_{1}^{\ast},z_{2}^{\ast}...z_{i}^{\ast}-\pi,...,z_{Ne}^{\ast})$ and
$\ \eta(z_{1}^{\ast},z_{2}^{\ast}...z_{i}^{\ast}-\pi\tau^{\ast},...,z_{Ne}%
^{\ast})$ in any of the particular coordinates $z_{i}^{\ast}.$

Let us argue below that  under those changes,  both of the functions $F_{L}$
and $F$ becomes themselves times a common factor function which cancels in the
ratio defining $\eta.$ This property implies that the proposed trial states
also satisfies the periodic boundary conditions in each of the particle
coordinates. \

\ \ For this purpose the transformations (\ref{per2}) allow to write
\begin{align*}
F(z_{1}^{\ast},z_{2}^{\ast}...z_{i}^{\ast}-\pi\tau^{\ast},...,z_{Ne}^{\ast})
&  =(-1)^{N_{e}}\exp\left(  -i\sum_{R}\left(  \frac{2\pi}{L_{1}}(z_{i}^{\ast
}-R^{\ast})-\pi\tau^{\ast}\right)  \right)  F(z_{1}^{\ast},z_{2}^{\ast
}...z_{i}^{\ast},...,z_{N_{e}}^{\ast}),\\
&  =(-1)^{N_{e}}\exp\left(  -i\frac{2\pi N_{e}}{L_{1}}z_{i}^{\ast}+\frac{2\pi
i}{L_{1}}\overline{R}^{\ast})+i\pi N_{e}\tau^{\ast}\right)  F(z_{1}^{\ast
},z_{2}^{\ast}...z_{i}^{\ast},...,z_{N_{e}}^{\ast}),
\end{align*}
where $\overline{R}=\sum_{R}R$ is the sum of all the complex coordinates of
the lattice $R$ which are in the first periodicity zone.

For the factor $F_{L}$ \ , on another hand, it follows%

\begin{align*}
F_{L}(z_{1}^{\ast},z_{2}^{\ast}...z_{i}^{\ast}-\pi\tau^{\ast},...,z_{Ne}%
^{\ast}) &  =(-1)^{N_{e}}\exp\left(  -i\sum_{j\neq i}\left(  \frac{2\pi}%
{L_{1}}(z_{i}^{\ast}-z_{j}^{\ast})-\pi\tau^{\ast}\right)  -i\frac{2\pi}{L_{1}%
}(Z^{\ast}-\overline{R})+i\pi\tau^{\ast}\right)  F_{L}(z_{1}^{\ast}%
,z_{2}^{\ast}...z_{i}^{\ast},...,z_{N_{e}}^{\ast})\\
&  =(-1)^{N_{e}}\exp\left(  -i\frac{2\pi N_{e}}{L_{1}}z_{i}^{\ast}+\frac{2\pi
i}{L_{1}}\overline{R}^{\ast})+i\pi N_{e}\tau^{\ast}\right)  F_{L}(z_{1}^{\ast
},z_{2}^{\ast}...z_{i}^{\ast},...,z_{N_{e}}^{\ast})
\end{align*}

Therefore, the similar way of transformation of the functions $F_{L}$ and $F$
implies the invariance of the factor $\eta$ under the translations in
$\ L_{1}\tau^{\ast}.$ \ The invariance under the shifts in $L_{1}\ $also is
directly obtained, thanks to the simplicity in the transformations
(\ref{per1}). The validity of these two  properties determines that the
proposed states also satisfy the periodicity conditions imposed over each of
the single particle variables. Thus, they are allowed states in the periodic
quantization of the electron system in the magnetic field .

\subsection{Case $\gamma=1$\ }

In this case the construction is simpler since the index $\sigma$ is absent.
The $HF$ single particle orbitals are $\varphi_{\mathbf{k}}^{(0)}(\mathbf{x})$
where the only quantum number is now the momentum%
\[
\Psi_{hf}^{(1)}=\Psi_{hf}^{(1)}[z_{1}^{\ast},z_{2}^{\ast},...z_{N_{e}}^{\ast
},z_{1},z_{2},...z_{N_{e}}]=Det[\varphi_{\mathbf{k}_{i}}^{(0)}(\mathbf{x}%
_{j})],
\]
in which the wave vectors $\mathbf{k}_{i}$ \ pertain to the lower now non
degenerate band defining the $HF$ state. Again, these momenta satisfy the
periodicity condition of the Haldane-Rezayi problem and their number in this
case is  equal to the number of electrons in the problem $N_{e}$. \ As shown
in the previous section the orbitals have the structure%
\begin{align*}
\varphi_{\mathbf{k}}^{(0)}(\mathbf{x}) &  =\prod_{R}\left\{  \left(
\vartheta_{1}(\frac{\pi}{L_{1}}(z^{\ast}-R^{\ast})|-\tau^{\ast})\right)
^{(q-1)}\right\}  \chi_{\mathbf{k}}^{(0)}(z^{\ast})\exp(k\ z^{\ast}%
+\frac{z^{\ast2}}{4r_{o}^{2}}-\frac{\mathbf{x}^{2}}{4r_{o}^{2}}),\\
&  =P(z^{\ast})\text{ }\chi_{\mathbf{k}}^{(0)}(z^{\ast})\exp(k\ z^{\ast}%
+\frac{z^{\ast2}}{4r_{o}^{2}}-\frac{\mathbf{x}^{2}}{4r_{o}^{2}}),\\
P(z^{\ast}) &  =\prod_{R}\left\{  \left(  \vartheta_{1}(\frac{\pi}{L_{1}%
}(z^{\ast}-R^{\ast})|-\tau^{\ast})\right)  ^{(q-1)}\right\}  ,\\
\chi_{\mathbf{k}}^{(0)}(z^{\ast}) &  \text{=}\prod_{R}\left\{  \vartheta
_{1}(\frac{\pi}{L_{1}}(z^{\ast}-R^{\ast}-C_{\mathbf{k}})|-\tau^{\ast
})\right\}  .
\end{align*}
where all \ the dependence of the momentum quantum numbers is contained in the
new function $\chi.$and the exponetial factor $\exp(k\ z^{\ast}).$

Similarly as in the last subsection the Slater determinant of the HF solution
can be represented as follows%

\begin{align*}
\Psi_{hf}^{(1)} &  =Det\text{ }[\varphi_{\mathbf{k}_{i}}^{(0)}(\mathbf{x}%
_{j})],\\
&  =(\prod_{k=1,2,..N_{e}}P(z_{k}^{\ast}))Det\text{ }[\chi_{\mathbf{k}_{i}%
}^{(0)}(z_{k}^{\ast})]\exp(\sum_{i=1,2...N_{e}}(k\ z_{i}^{\ast}+\frac
{z_{i}^{\ast2}}{4r_{o}^{2}}-\frac{\mathbf{x}_{i}^{2}}{4r_{o}^{2}})),\\
&  =\Phi(z_{1}^{\ast},z_{2}^{\ast},...z_{N_{e}}^{\ast})D[z_{1}^{\ast}%
,z_{2}^{\ast},...z_{N_{e}}^{\ast}]\exp(\sum_{i=1,2...N_{e}}(\frac{z_{i}%
^{\ast2}}{4r_{o}^{2}}-\frac{\mathbf{x}_{i}^{2}}{4r_{o}^{2}})),\\
\Phi(z_{1}^{\ast},z_{2}^{\ast},...z_{N_{e}}^{\ast}) &  =\prod_{R}\left\{
\left(  \vartheta_{1}(\frac{\pi}{L_{1}}(z^{\ast}-R^{\ast})|-\tau^{\ast
})\right)  ^{q-1}\right\}  ,\\
D[z_{1}^{\ast},z_{2}^{\ast},...z_{N_{e}}^{\ast}] &  =\exp(\sum_{i=1,2...N_{e}%
}k\ z_{i}^{\ast})\text{ }Det\text{ }[\chi_{(\mathbf{k}_{i}\mathbf{)}}%
^{(0)}(z_{k}^{\ast})].
\end{align*}
\ \

Again, the new factors $\Phi$ \ when considered as function of any of the
electron coordinates $z^{\ast},$ have $(q-1)N_{e}$ zeroes. \ Since the
determinant $Det$ $[\chi_{\mathbf{k}_{i}}^{(0)}(z_{j}^{\ast})]$ \ also shows a
number \ $N_{e}$ of \ zeroes, the total number of vanishing poits  of
$\ \Psi_{hf}^{(1)}$ (with respect to any particular variable) is again equal
to  the number of flux quanta. \

Thus, following the same steps as in the last section, in this case, the
variational states for this case are proposed in the form
\begin{align*}
\Psi_{N}^{(1)} &  ==\Phi_{L}(z_{1}^{\ast},z_{2}^{\ast},...z_{N_{e}}^{\ast
})D^{(1)}[z_{1}^{\ast},z_{2}^{\ast},...z_{N_{e}}^{\ast}]\exp(\sum
_{i=1,2...N_{e}}(\frac{z_{i}^{\ast2}}{4r_{o}^{2}}-\frac{\mathbf{x}_{i}^{2}%
}{4r_{o}^{2}}))\\
\Phi_{L}(z_{1}^{\ast},z_{2}^{\ast},...z_{N_{e}}^{\ast}) &  =\left\{
\vartheta_{1}(\frac{\pi}{L_{1}}(Z^{\ast}-\overline{R}^{\ast})|-\tau^{\ast
})\right\}  ^{q-1}\prod_{\substack{i<j\\i,j=1,2...N_{e}}}\left\{
\vartheta_{1}(\frac{\pi}{L_{1}}(z_{i}^{\ast}-z_{j}^{\ast})|-\tau^{\ast
})\right\}  ^{q-1},\\
D^{(1)}(z_{1}^{\ast},z_{2}^{\ast},...z_{N_{e}}^{\ast}) &  =\exp(\sum
_{i=1,2...N_{e}}k\ z_{i}^{\ast})\text{ }Det\text{ }[\chi_{\mathbf{k}_{i}%
}^{(0)}(z_{k}^{\ast})].
\end{align*}

In an almost identical way, as in the previous section it can be argued that
these functions \ satisfy the periodic boundary conditions.

\section{Summary}

The Haldane-Rezayi procedure for describing periodic boundary
conditions in presence of a magnetic field is employed for
constructing new variational states which are expected to optimize
short range correlations in a similar form as the Laughlin states.
However, their construction also include crystalline correlations
which were shown to optimize the mean field HF energies. \ Two
classes \ of states are proposed which described by the number of
electrons per unit cell $\gamma$ which defines the associated HF
solution. The cases $\gamma=1,1/2$ are considered for any filling
factor of the forms $1/q$. They detrmine two classes of
alternative states: one showing $q$ flux quanta per periodicity
cell and the other having $q/2$ quanta per cell. \ \ The states
are shown to satify \ the periodic boundary conditions. Also, as
it should be,  the wave functions when  considered as  functions
of any of the identical particle coordinates, they show a number
of zeroes which is equal \ to the number of flux quanta piercing
the periodicity area of the system \cite{halrez}. For the
considered cases of filling factors $\nu=1/q$ the  number of flux
quanta is $qN_{e}$, where $N_{e}$ is the number of electrons in
the sample. \

The main elements of the proposal can be schematically resumed as follows. Let
us consider in particular the HF Slater determinant of single particle
orbitals described Ref. [\onlinecite{ccpm}] for the case $\gamma=1$. In these
determinant $(q-1)N_{e}$ of its zeros as a function of the coordinates of any
given particle are fixed a the points of the periodicity lattice. These zeroes
appear in a factor function $\Phi$ which is completely independent of the
quantum numbers of the filled orbitals. Then, this basic property allows to
transform the mean field Slater determinant by substituting the factor
function $\Phi$, by a factor function $\Phi_{L}$ having the same number of
zeroes that $\Phi$ \ but only depending on the differences of the coordinates
in the usual Jastrow way. This procedure suggests a possibility for a lowering
of the energies per particle, by reducing short range Coulomb repulsion. This
is related with the fact that the zeroes of the proposed states,  when the
coordinates of two particles equalize,  is of the same order $q$ as in the
Laughlin states. It should be noticed that the $N_{e}$ zeroes associated to
the functions $D^{(1)}$ and $D^{(\frac{1}{2})}$\ have spatial locations
depending on the quantum number $\mathbf{k}$, which label each occupied
orbital in the factor having the form of an Slater determinant.  \ It could be
the case that the presence of remaining crystalline mean field information in
the determinantal factors $D^{(1)}$ and $D^{(\frac{1}{2})}$  could lower the
energy per particle of the Laughlin anstaz. \ The evaluation of the
correlation energy for these states is under consideration. We hope that this
proposal can stimulate further research directed to check whether the
corrections to the mean field approach early introduced by one of the authors
(F.C), could furnish a general theory for the FQHE, showing the relevant
Laughlin or Jain procedures as closely approximate limits.

\begin{acknowledgments}
We acknowledge partial support from Fondecyt, Grants 1060650 and
7020829, the Catholic University of Chile and the Caribbean
Network on Quantum Mechanics, Particles and Fields of the ICTP
Office of External Activities (OEA).
\end{acknowledgments}

\end{document}